\documentclass[journal]{IEEEtran}
\usepackage{graphicx}

\begin{document}
\title{Performance of the micro-TPC Reconstruction for GEM Detectors at High Rate}  

\author{L.~Lavezzi$^{*,a,f}$, M.~Alexeev$^f$, A.~Amoroso$^{f,l}$, R.~Baldini Ferroli$^{a,c}$, M.~Bertani$^c$, D.~Bettoni$^b$, F.~Bianchi$^{f,l}$, A.~Calcaterra$^c$, N.~Canale$^b$, M.~Capodiferro$^{c,e}$, V.~Carassiti$^b$, S.~Cerioni$^c$, JY.~Chai$^{a,f,h}$, S.~Chiozzi$^b$, G.~Cibinetto$^b$, F.~Cossio$^{f,h}$, A.~Cotta Ramusino$^b$, F.~De Mori$^{f,l}$, M.~Destefanis$^{f,l}$, J.~Dong$^c$, F.~Evangelisti$^b$, R.~Farinelli$^{b,i}$, L.~Fava$^f$, G.~Felici$^c$, E.~Fioravanti$^b$, I.~Garzia$^{b,i}$, M.~Gatta$^c$, M.~Greco$^{f,l}$, CY.~Leng$^{a,f,h}$, H.~Li$^{a,f}$, M.~Maggiora$^{f,l}$, R.~Malaguti$^b$, S.~Marcello$^{f,l}$, M.~Melchiorri$^b$, G.~Mezzadri$^{b,i}$, M.~Mignone$^f$, G.~Morello$^c$, S.~Pacetti$^{d,k}$, P.~Patteri$^c$, J.~Pellegrino$^{f,l}$, A.~Pelosi$^{c,e}$, A.~Rivetti$^f$, M.~D.~Rolo$^f$, M.~Savri\'e$^{b,i}$, M.~Scodeggio$^{b,i}$, E.~Soldani$^c$, S.~Sosio$^{f,l}$, S.~Spataro$^{f,l}$, E.~Tskhadadze$^{c,g}$, S.~Verma$^i$, R.~Wheadon$^f$, L.~Yan$^f$ \\

  \thanks{$*$ speaker, e-mail: lia.lavezzi@to.infn.it}
  \thanks{$^a$ Institute of High Energy Physics, Beijing, China}
  \thanks{$^b$ INFN, Sezione di Ferrara, Italy}
  \thanks{$^c$ INFN, Laboratori Nazionali di Frascati, Italy}
  \thanks{$^d$ INFN, Sezione di Perugia, Italy}
  \thanks{$^e$ INFN, Sezione di Roma, Italy}
  \thanks{$^f$ INFN, Sezione di Torino, Italy}
  \thanks{$^g$ Joint Institute for Nuclear Research (JINR), Dubna, Russia}
  \thanks{$^h$ Politecnico di Torino, Italy}
  \thanks{$^i$ Universit\`a di Ferrara, Italy}
  \thanks{$^k$ Universit\`a di Perugia, Italy}
  \thanks{$^l$ Universit\`a di Torino, Italy}
  for the CGEM-IT Group
}


\maketitle
\pagestyle{empty}
\thispagestyle{empty}

\begin{abstract}
  Gas detectors are one of the pillars of the research in fundamental physics. Since several years, a new concept of detectors, called Micro Pattern Gas Detectors (MPGD), allows to overcome many of the problems of other types of commonly used detectors, like drift chambers and microstrip detectors, reducing the discharge rate and increasing the radiation tolerance. \\
  Among these, one of the most commonly used is the Gas Electron Multiplier (GEM). GEMs have become an important reality for fundamental physics detectors. Commonly deployed as fast timing detectors and triggers, due to their fast response, high rate capability and high radiation hardness, they can also be used as trackers. \\
  The readout scheme is one of the most important features in tracking technology. The center of gravity technique allows to overcome the limit of the digital pads, whose spatial resolution is constrained by the pitch dimension. The presence of a high external magnetic field can distort the electronic cloud and affect the spatial resolution. The micro-TPC ($\mu-$TPC) reconstruction method allows to reconstruct the three dimensional particle position as in a traditional Time Projection Chamber, but within a drift gap of a few millimeters. This method brings these detectors into a new perspective for what concerns the spatial resolution in strong magnetic field. \\
 In this report, the basis of this new technique will be shown and it will be compared to the traditional center of gravity. The results of a series of test beam performed with $10 \times 10$ cm$^2$ planar prototypes in magnetic field will also be presented. \\
 This is one of the first implementations of this technique for GEM detectors in magnetic field and allows to reach unprecedented performance for gas detectors, up to a limit of $120$ $\mu$m at $1$ T, one of the world's best results for MPGDs in strong magnetic field. The $\mu-$TPC reconstruction has been recently tested at very high rates in a test beam at the MAMI facility; preliminary results of the test will be presented.
\end{abstract}


\section{Introduction}

\begin{figure}[!ht]
\centering
\includegraphics[width=0.8\columnwidth]{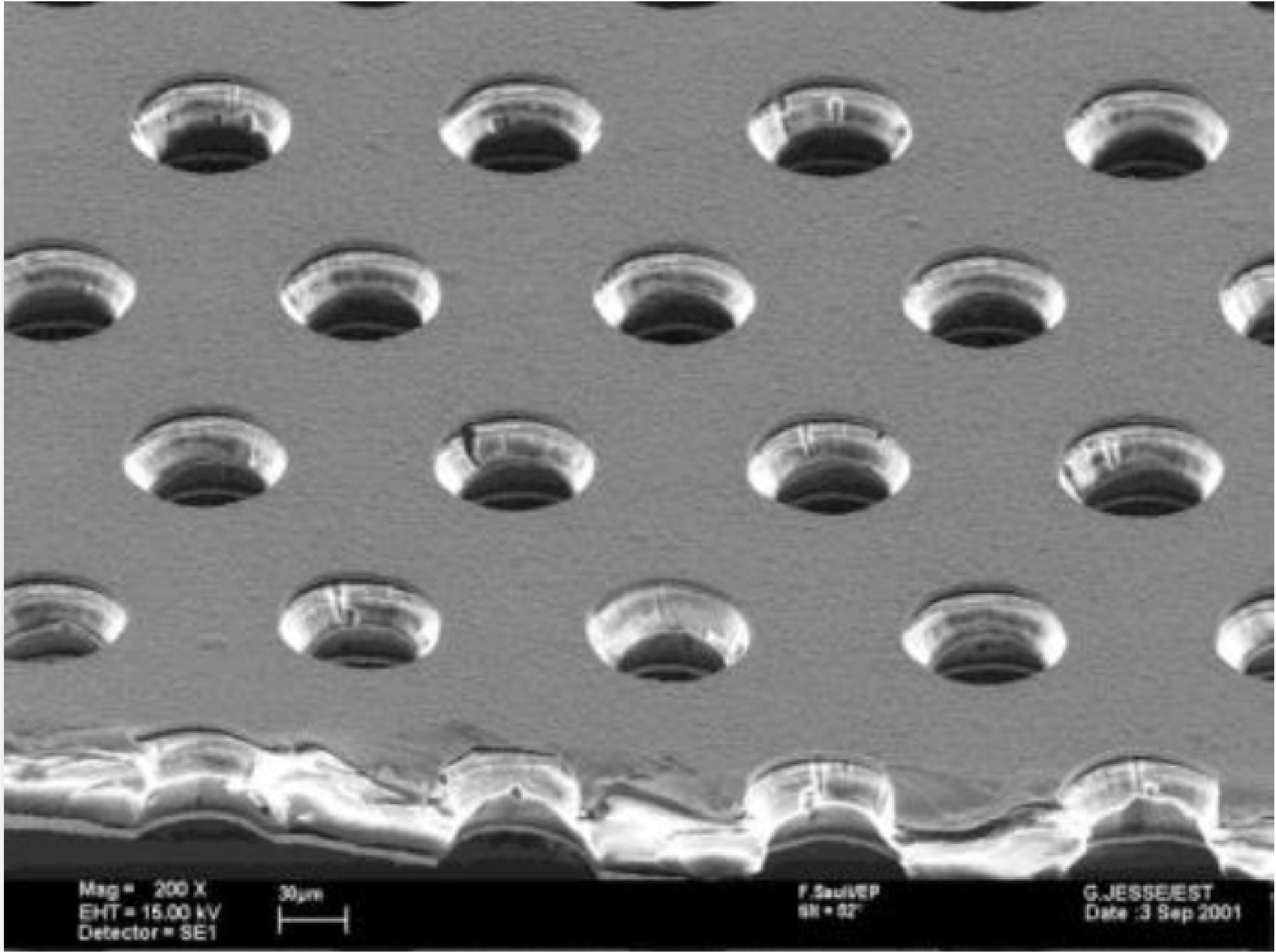}
\caption{Detail of a GEM foil \cite{sauli}.} \label{fig:gem}
\end{figure}

\IEEEPARstart{H}{igh} energy physics research requires a constant improvement in the machine performance. For example, the increasing accelerator luminosity, which grants the possibility to acquire big samples of data, forces the detectors to keep up with an always bigger particle rate. This reflects on the need to both choose detectors with a good capability to undergo strong radiation doses without relevant aging and to develop new reconstruction methods, able to cope with the {\it crowded} environment. \\
In $1997$ F. Sauli \cite{sauli} invented the Gas Electron Multiplier (GEM) to allow gas-based trackers to work under higher particle rates \cite{pdg}. \\
In a standard gas-based tracker, the charged particle ionizes the gas producing electrons and positive ions. The electrons follow the electric drift field lines to a region where the electric field becomes so intense they undergo avalanche multiplication. The obtained amount of electrons is sufficient to induce a signal on the readout. In standard trackers the high electric field is generated by wires, but this creates a problem of discharge already at $10^3$ Hz mm$^{-2}$. \\
A more robust way to obtain electron multiplication is the GEM: it consists of a thin ($\sim 50$ $\mu$m) polymeric foil, covered on both sides by two thinner ($\sim 3$ $\mu$m) layers of copper. The foil is pierced with thousands of double-conical holes, with an inner diameter of $50$ $\mu$m (see fig.\ref{fig:gem}).
A voltage of some hundreds of Volts is applied between the copper layers and due to the tiny dimensions of the holes it creates an electric field of some kV/cm inside them. When the electrons resulting from the ionization of the gas move along the drift field lines and enter the holes, they meet an electric field intense enough to produce avalanche multiplication with a gain of some $10^4$. This makes GEM-based detectors more rate tolerant than wire-based ones. The effect is even more emphasized when more GEM foils are placed in series, instead of just one \cite{bachman}.

\section{The reconstruction methods}
\begin{figure}[!ht]
\centering
\includegraphics[width=0.7\columnwidth]{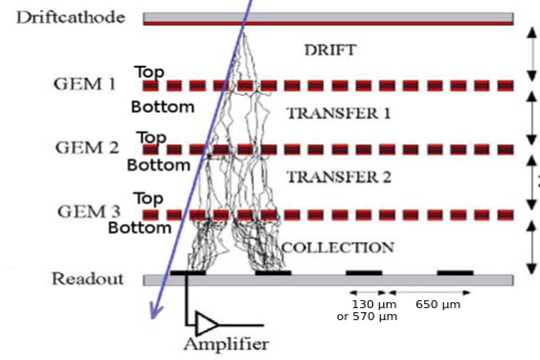}
\caption{The different sections of a triple-GEM.} \label{fig:triple_gem}
\end{figure}

Being tracking detectors, the position reconstruction is their primary goal. Two algorithms are currently available to reconstruct the particle position: the center of gravity method, commonly called {\it charge centroid} (CC) and the micro-TPC method ($\mu-$TPC). The choice between them depends strongly on the shape of the charge distribution on the readout plane. \\
The standard layout of a triple-GEM detector is showed in fig.\ref{fig:triple_gem}.
It consists of:
\begin{itemize}
\item a cathode, at negative potential;
\item a drift gap, where the ionization happens;
\item a set of three GEM foils, with (transfer) gas gaps between them, to produce the multiplication;
\item an induction gap, where the induction of the signal on the final electrode, the anode, begins;
\item an anode, at ground, where the strips of the readout are placed and the signal is collected. Usually two views are available, for 2D reconstruction on the plane.
\end{itemize}
If the anode plane is digitally readout, then the {\it on/off} information is the only one available from the strips and the pitch dominates the position resolution. If an analog readout is applied, the additional value of the charge deposited on each strip is available, allowing a more refined position determination. Moreover, if also the time of arrival of the signal is recorded a new degree of freedom opens up in the position reconstruction, as will be shown in the following. \\
The shape of the charge distribution on the anode is determined by the physical effects which come into play along the electron path from the primary ionization points to the readout plane. The most important effects are the diffusion and the possible presence of the Lorentz force. \\
The former is due to the motion inside the gas: the multiple scattering enlarges the electronic cloud and spreads it over more than one strip on the readout plane. \\
The second effect is present if a magnetic field is applied, usually orthogonal to the electric drift field. The electron trajectories are bent and the charge distribution at the anode is not only enlarged but its shape goes also far from the Gaussian shape and is no longer parametrizable.
\subsection{Charge centroid}
The charge centroid consists in the weighted average of the firing strip positions, the weights being the charge measured on each strip (eq.\ref{eq:cc}).
\begin{equation} 
  x = \frac{\sum\limits_i x_i q_i}{\sum\limits_i q_i} \label{eq:cc}
\end{equation}
It is the simplest reconstruction possible by knowing the strip positions and the charge values and is well performing when the charge distribution shape is Gaussian. For non Gaussian shapes it does not provide a good position resolution anymore. Inclined incident tracks at big angles and/or the presence of a strong magnetic field may create a situation where the CC method cannot guarantee a good spatial resolution. In these cases, when the CC fails, another method must be adopted: the $\mu-$TPC.
\subsection{$\mu-$TPC mode}
\begin{figure}[!ht]
\centering
\includegraphics[width=0.8\columnwidth]{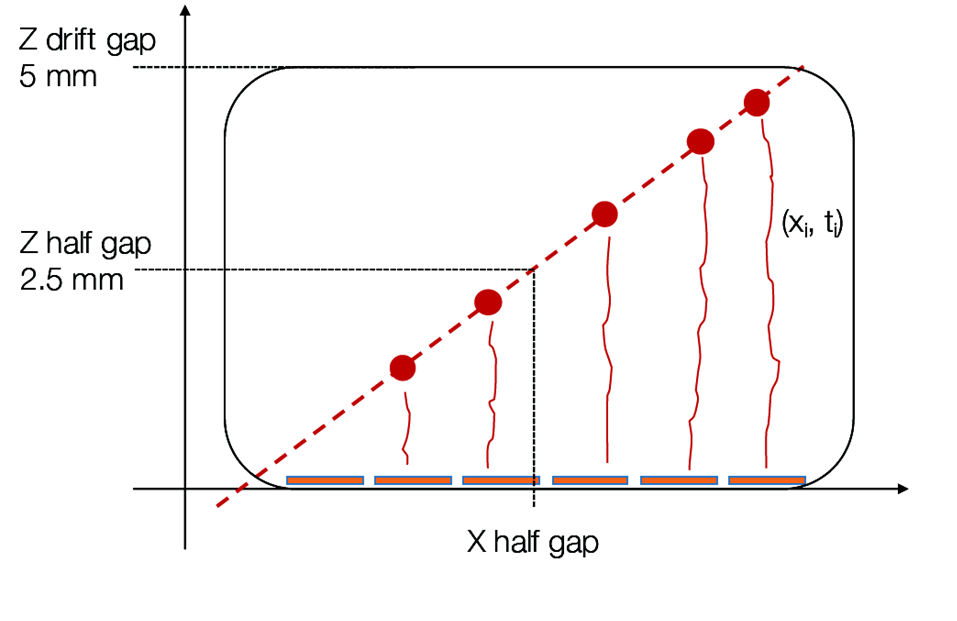}
\caption{Sketch of the track reconstruction inside the drift gap with the $\mu-$TPC.} \label{fig:utpc}
\end{figure}

\begin{figure}[!ht]
\centering
\includegraphics[width=0.9\columnwidth]{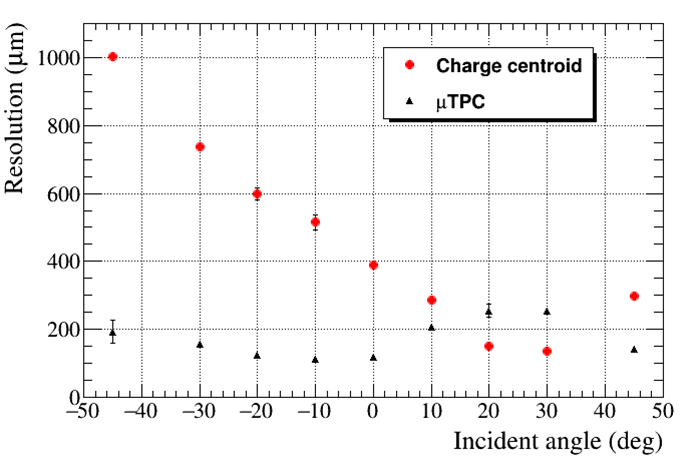}
\caption{Spatial resolution of the CC and $\mu-$TPC cluster reconstruction {\it vs} the incident angle of the track for Ar$:$iC$_4$H$_{10}$ ($90:10$) gas mixture at B $= 1$ T. Results obtained with a drift gap of $5$ mm, a drift field of $1.5$ kV/cm and a gain of $9000$ \cite{ric_ieee16}.} \label{fig:resol}
\end{figure}

This method was firstly introduced in ATLAS, for the Micromegas detector \cite{utpc}. As the name says, the idea behind it is to use the GEM drift gap as a {\it micro Time Projection Chamber}. By measuring the time of arrival of the signal on each strip and by knowing the drift velocity in the specific gas under the working conditions it is possible to calculate the position of the primary ionization point. \\
In fig.\ref{fig:utpc}
the $\mu-$TPC concept is sketched. Once a cluster is found, from the $x$ position of the strip on the anode and the $z$ position of the primary ionization in the drift gap, couples of $(x, z)$ coordinates and $(dx,dz)$ errors are assigned to each strip and a fit with a straight line is performed. The $dx$ errors basically account for the uncertainty of the hit in the finite strip pitch plus a weight depending on the charge on the strip; the error $dz$ results from the propagation of the time measurement error. The best position measurement (eq.\ref{eq:utpc}) corresponds to the track fit at half-gap, where the interpolated position estimate minimizes the error.
\begin{equation} \label{eq:utpc}
x = \frac{\frac{gap}{2} - b}{a}
\end{equation}
The whole procedure is possible if the time resolution of the detector is good enough to resolve the arrival times of the electron avalanche on different strips, and with a highly segmented readout plane. \\
The $\mu-$TPC clustering method has been initially tested with inclined tracks and magnetic field up to $1$ T. Data samples with chambers at $10^\circ$, $20^\circ$, $30^\circ$ and $45^\circ$ w.r.t. the beam direction have been collected. A data driven correction procedure, based on the identification of the strip signals by induced charge (based on the time information and charge ratios) and subsequent weighting or suppression of the first and/or last strips in the cluster has been implemented. Fig.\ref{fig:resol}
shows the resolution for a $5$ mm gap prototype with Ar$:$iC$_4$H$_{10}$ ($90:10$) gas mixture as a function of the incident angle, along with the comparison with the CC performance: spatial resolution around $130$ $\mu$m is achievable for all the impact angles by combining the CC and the $\mu-$TPC results. \\
This is the first implementation of the $\mu-$TPC algorithm for a GEM detector in a strong magnetic field. A more detailed description of these results can be found in \cite{ric_ieee16}.

\section{High rate test beam}
As already anticipated, the behavior of the detector experiencing high particle rates changes. \\
In every gas detector, when one ionization happens, the electron and the positive ion drift along the electric field lines; the electron drift velocity is high and it produces a fast signal, while the ion mobility is lower and at high rates there might be an accumulation of positive charge inside some areas of the detector. This space charge issue may create a distortion in the electric field lines and a consequent reduction of the gain. This can lead to a degradation of the spatial resolution and eventually to aging. The limit for wire detectors is known to be around $10^3$ Hz mm$^{-2}$ but the GEMs can resist to much higher rates \cite{pdg}.

\subsection{The environment and the setup}
\begin{figure}[!ht]
\centering
\includegraphics[width=0.8\columnwidth]{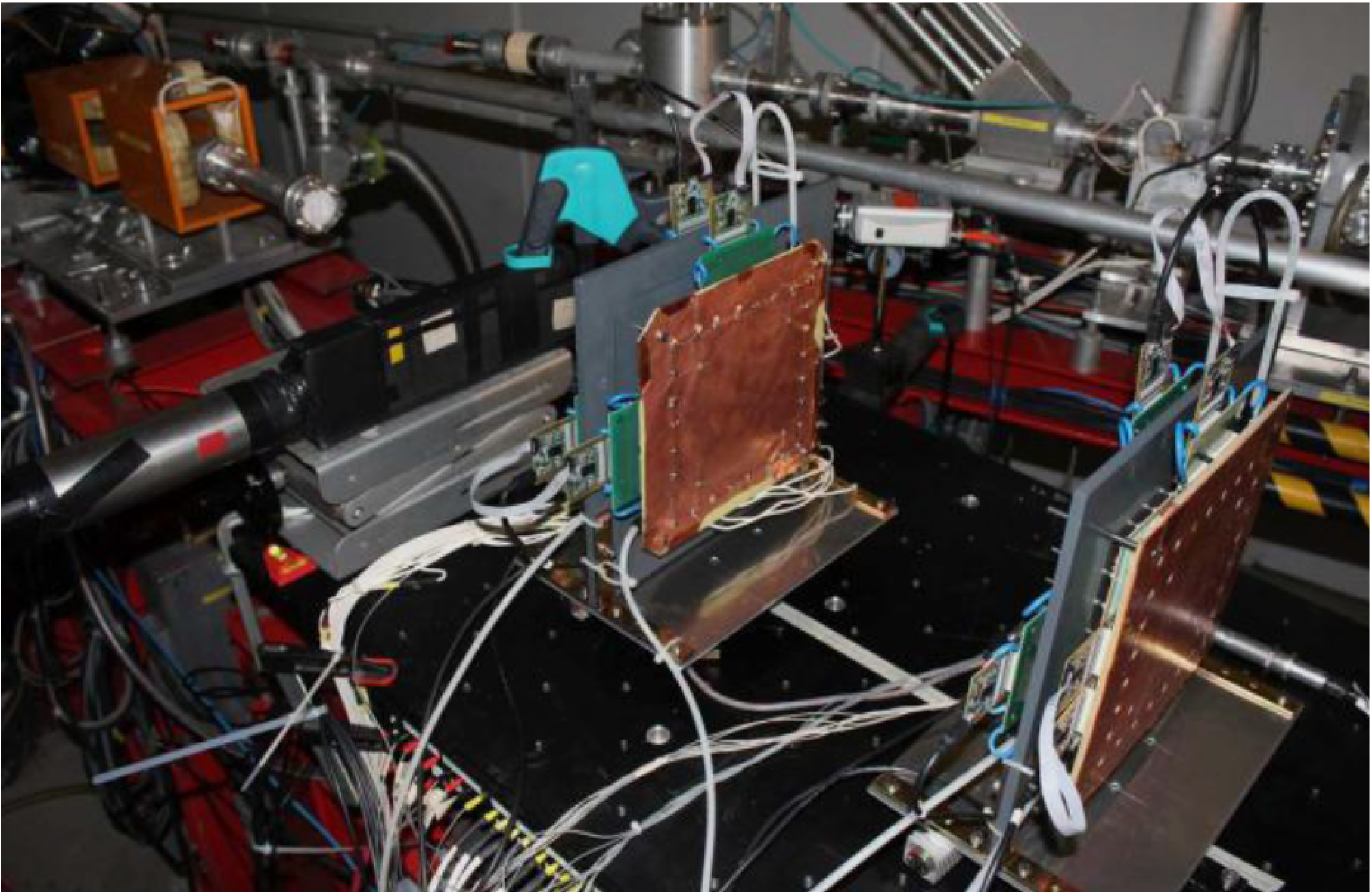}
\caption{Setup installed at MAMI facility.} \label{fig:mamisetup}
\end{figure}

The effect of high rate on the $\mu-$TPC performance has been recently studied with a test beam at the MAinz MIcrotron, MAMI \cite{mami} facility, in Mainz, Germany. This test was necessary since the detector must not only be able to sustain high rates without damage, but also keep a level of performance unaltered by the challenging environment. \\
The setup (shown in fig.\ref{fig:mamisetup}) consisted in four triple-GEM planar chambers $10 \times 10$ cm$^2$ with a $5$ mm drift gap. The tested gas mixtures were Ar$:$iC$_4$H$_{10}$ ($90:10$) and Ar$:$CO$_2$ ($70:30$) without magnetic field. The chambers could rotate and the $\mu-$TPC studies were performed at an angle of $30^\circ$ w.r.t the beam direction: the angle is necessary since the $\mu-$TPC is not applicable at $0^\circ$. The electron beam had a size of a few mm and could reach high rates. \\

\subsection{The results}
\begin{figure}[!ht]
\centering
\includegraphics[width=0.9\columnwidth]{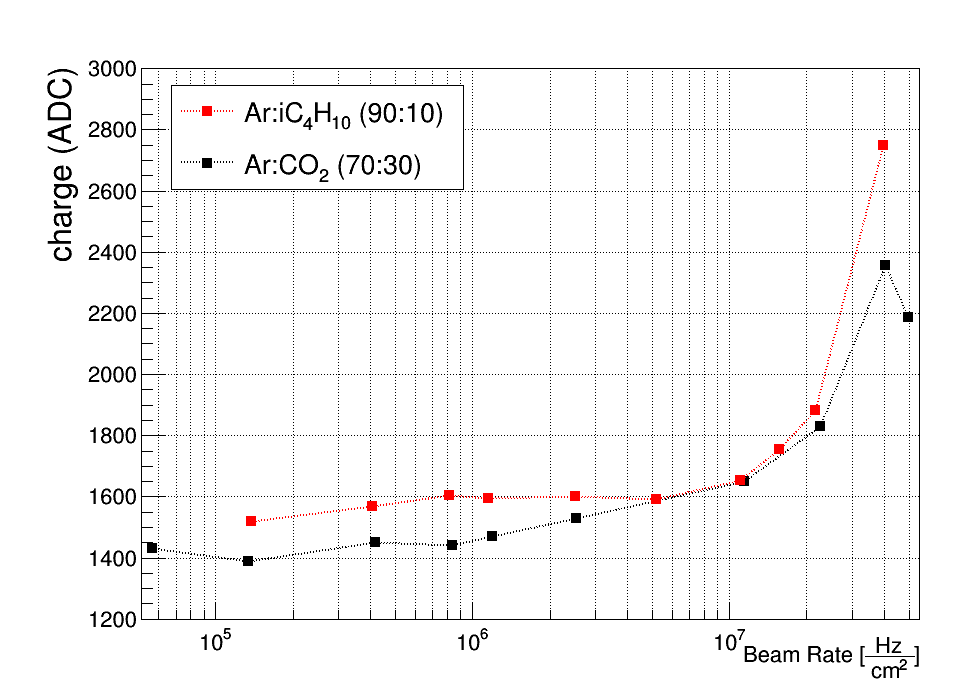}
\caption{Charge {\it vs} rate.} \label{fig:charge}
\end{figure}

\begin{figure}[!ht]
\centering
\includegraphics[width=0.8\columnwidth]{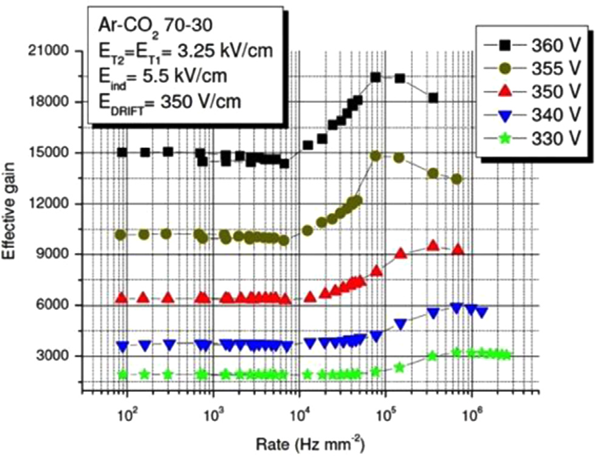}
\caption{Gain {\it vs} rate from \cite{sauli2016}.} \label{fig:gain}
\end{figure}

\begin{figure}[!ht]
\centering
\includegraphics[width=0.9\columnwidth]{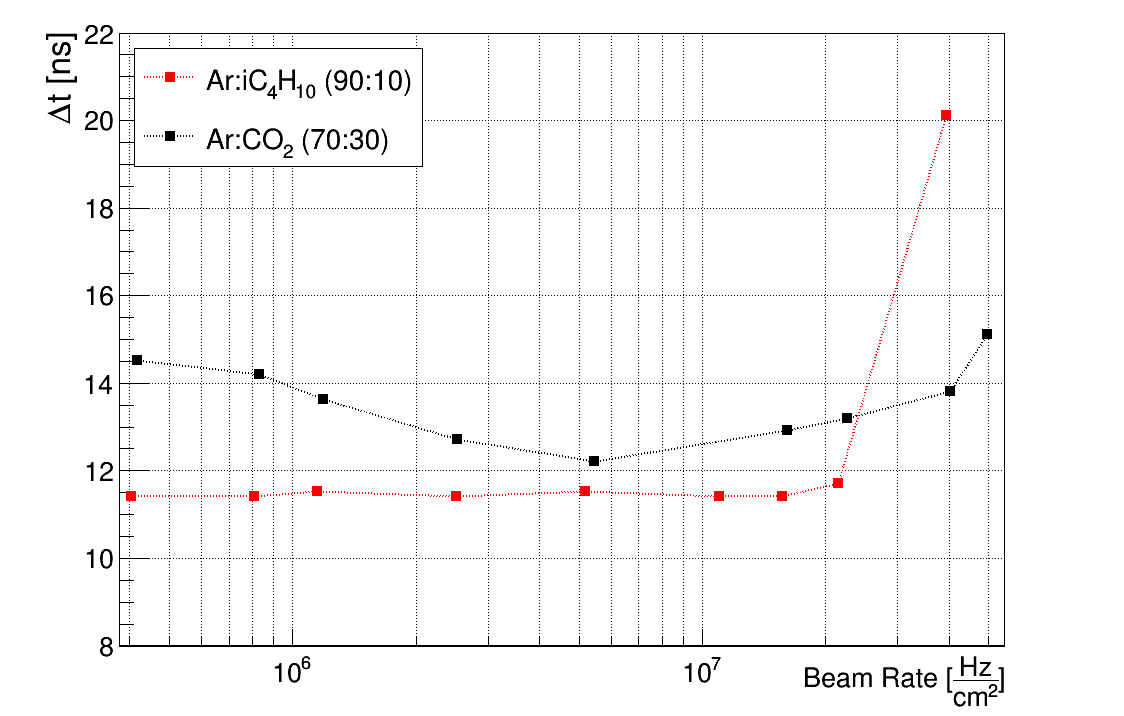}
\caption{$\Delta t$ {\it vs} rate.} \label{fig:time}
\end{figure}

\begin{figure}[!ht]
\centering
\includegraphics[width=0.9\columnwidth]{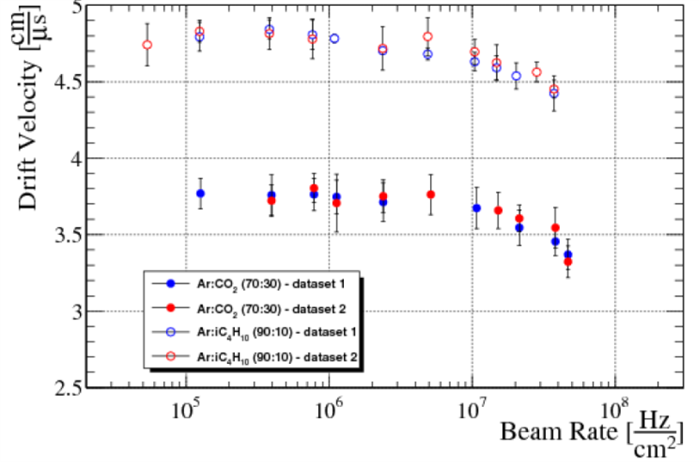}
\caption{Drift velocity {\it vs} rate.} \label{fig:velocity}
\end{figure}

A key factor for GEM reliability is a stable gain value. Moreover, the main parameters the $\mu-$TPC depends on are the drift velocity and the time resolution. For these reasons, the variation with increasing particle rate of these variables has been studied. \\
\, \\
Fig.\ref{fig:charge} shows that the cluster mean charge is constant up to $10^6$ Hz cm$^{-2}$, then it increases up to $10^7$ Hz cm$^{-2}$ and eventually it drops. This behavior has a big resemblance to the one shown for the gain in Sauli's recent review on GEMs \cite{sauli2016}, that we report here for simplicity (fig.\ref{fig:gain}). When we {\it scale} the charge {\it vs} rate plot to a gain {\it vs} rate one, the resulting behavior is compatible with the one shown in fig.\ref{fig:gain}, even though a direct comparison of the values is not possible due to the different electrical settings. The gain is stable up to $10^6$ Hz cm$^{-2}$, increases up to $10^7$ Hz cm$^{-2}$ and drops afterwards. \\
An explanation for this peculiar behavior is given in \cite{thuiner}: the space charge due to the positive ions modifies the electric field and increases the transparency of the GEM. The transparency is defined as the collection efficiency multiplied by the extraction efficiency, i.e. the probability that an electron drifts inside the GEM hole multiplied, after the avalanche, by the fraction of the electrons which exit from it. Usually the electric fields in the various gas gaps, especially the transfer gaps between the GEMs, must be conveniently optimized to find a compromise between the extraction efficiency of the previous GEM and the collection efficiency of the following one. The positive charge accumulation due to the high rates modifies the electric field in these regions in such a way that both these efficiencies are enhanced and the full triple-GEM becomes {\it more transparent}. This increases the effective gain. \\
This situation is transitory and when the space charge is too high the gain suddenly falls to a lower value. \\
\, \\
The time resolution can be evaluated by distributing the time difference measured by two adjacent strips for the same event. This $\Delta t$ contains not only the resolution of the detector, but also the effect of the intrinsic time resolution of the electronics. In this test beam, the data were collected through the APV-25 ASIC \cite{apv}, which samples the charge every $25$ ns. By deconvoluting the $\Delta t$ from the APV-25 contribution, the obtained resolution results in $8.4$ ns for the Ar$:$iC$_4$H$_{10}$ mixture. \\
The $\Delta t$ {\it vs} rate is shown in fig.\ref{fig:time}: it starts worsening only after $10^7$ Hz cm$^{-2}$. \\
\, \\
The last evaluated parameter was the drift velocity. It can be extracted from the data by plotting all the measured times and computing the difference between the leading and the falling edges of the obtained distribution. Its behavior as a function of the rate (fig.\ref{fig:velocity}) shows that, again, there is a relevant change only after $10^7$ Hz cm$^{-2}$. At higher rates the electrons are slower and this is expected to affect the $\mu-$TPC performance. \\

\section{Conclusion}
The test at MAMI confirms that no relevant changes occur in the parameters that influence the $\mu-$TPC reconstruction resolution up to $10^7$ Hz cm$^{-2}$ and this gives good expectations on the $\mu-$TPC behavior up to this particle rate. More studies and the actual reconstruction of the collected data with the $\mu-$TPC mode are necessary to certify its applicability at these rate levels. This, however, was the first test on the limits of the $\mu-$TPC at high rates. \\
As additional result, we observed that the gain showed the peculiar behavior seen in previous tests.

\section*{Acknowledgment}
The authors wish to thank Werner Lauth (MAMI) for his help in the test beam realization, as well as Giovanni Bencivenni (LNF) and Eraldo Oliveri (CERN) for the fruitful discussion on the results. \\
The research leading to these results has been performed within the BESIIICGEM Project, funded by the European Commission in the call H2020-MSCA-RISE-2014.

\end{document}